\documentclass[usenatbib,usegraphicx]{mn2e}
\usepackage{timesfonts}
\usepackage{journals}
\begin{document}

\title[Understanding white dwarf interiors]{A deeper understanding of white dwarf interiors}

\author[T.~S. Metcalfe]{Travis S. Metcalfe\thanks{NSF Astronomy \& Astrophysics Postdoctoral Fellow}\\
High Altitude Observatory, National Center for Atmospheric Research, P.O. Box 3000, Boulder, CO 80307-3000 USA}

\maketitle

\begin{abstract} 
A detailed record of the physical processes that operate during 
post-main-sequence evolution is contained in the internal chemical 
structure of white dwarfs. Global pulsations allow us to probe the stellar 
interior through asteroseismology, revealing the signatures of prior 
nuclear burning, mixing, and diffusion in these stars. I review the rapid 
evolution of structural models for helium-atmosphere variable (DBV) white 
dwarfs over the past five years, and I present a new series of model-fits 
using recent observations to illustrate the relative importance of various 
interior structures. By incorporating physically motivated C/O profiles 
into double-layered envelope models for the first time, I finally identify 
an optimal asteroseismic model that agrees with both diffusion theory and 
the expected nuclear burning history of the progenitor. I discuss the 
implications of this fundamental result, and I evaluate the prospects for 
continued progress in the future.
\end{abstract}



\section{INTRODUCTION}

The internal structure of a white dwarf star contains a detailed record of 
the physical processes that operate during post-main-sequence evolution. 
The composition of the deep interior is determined by the nuclear reaction 
rates during core helium burning in the red giant progenitor, while the 
chemical profile in the outer core is set by helium shell burning and 
various mixing processes including convective overshoot. More than 99\% of 
the mass in a typical white dwarf is contained in the C/O core, which is 
embedded in an envelope of C, He, and H with traces of heavier elements. 
Near the end of post-AGB evolution the white dwarf envelope can become 
uniformly mixed during a very late thermal pulse, which burns off most of 
the residual H and creates a born-again AGB star \citep{ibe83}. As this 
newly formed H-deficient star begins to descend the white dwarf cooling 
track, the high surface gravity ($\log~g\sim8$) coupled with chemical and 
thermal diffusion leads to compositional stratification. The lighter 
elements float to the surface, slowly building a nearly pure surface He 
layer above the remainder of the envelope as the star cools.

Asteroseismology is the only observational method available to probe the 
interior structure of white dwarfs and calibrate theories of nuclear 
burning, mixing, and diffusion in these stars. The helium-atmosphere (DB) 
white dwarfs are particularly suitable for asteroseismic study, since they 
are observed to pulsate in a range of effective temperatures between 
22,400 and 28,400~K \citep{bea99}. The mass of the pure He surface layer 
is predicted to grow by roughly an order of magnitude within this range, 
so an asteroseismic calibration of diffusion theory is possible by 
studying DB variables (DBVs) at various temperatures 
\citep{dk95,fb02,ac04,met05}. Unlike the hotter class of variable white 
dwarfs (the PG~1159 stars) DBVs do not experience significant 
gravitational contraction, which simplifies the asteroseismic analysis 
since the mechanical and thermal structure are decoupled \citep{win04}. 
The cooler hydrogen-atmosphere variables (DAVs) also enjoy this advantage, 
but their cores are significantly more degenerate---reducing the influence 
of the deep interior structure on their pulsation periods.

Despite their extraordinary potential, only two DBV white dwarfs have been 
sufficiently characterized to allow detailed asteroseismic study. The 
brightest member of the class, GD~358, has been the target of three 
multi-site observing campaigns of the Whole Earth Telescope 
\citep[WET;][]{nat90,win94,vui00,kep03}. These data revealed a series of 
11 dipole modes \citep[$\ell$=1,~$m$=0;][]{kot03} of consecutive radial 
overtone \citep[$k$=8-18;][]{bw94} that could be used for model-fitting. 
Prior to the Sloan Digital Sky Survey \citep[SDSS;][]{nit05} the faintest 
member of the class was CBS~114, which was first subjected to detailed 
study by \citet{hmw02}. A recent dual-site campaign on this star by 
\citet{met05} documented a total of 11 dipole modes with radial overtones 
ranging from $k$=8 to $k$=20.

In the following section, I review the recent development of detailed 
structural models for DBV white dwarfs with an emphasis on the published 
models for GD~358 over the past five years. I then present a series of new 
model-fits to the latest observations of CBS~114 in Section~\ref{SEC3} to 
illustrate the relative importance of the various interior structures 
expected from stellar evolution theory. The final fit in this series 
represents the first successful application of a model containing a 
complete, physically motivated description of the white dwarf interior. 
Finally in Section~\ref{SEC4}, I discuss the implications of this result 
for theories of diffusion and nuclear burning, and I evaluate the 
prospects for future progress.


\section{MODEL DEVELOPMENT}

Although detailed observations of the DBV white dwarf GD~358 have been 
available since 1994, the realization of their full potential had to wait 
for the large-scale exploration of models made possible by fast, 
inexpensive computers. The original analysis by \citet{bw94} attempted to 
match the observed pulsation periods using a small grid of models around 
an initial guess based on general scaling arguments and analytical 
relations developed by \citet{kaw90}, \citet{kw90}, \citet{bra92}, and 
\citet{bww93}. The main physical parameters that were adjusted to fit the 
models to the data included the stellar mass ($M_*$), the effective 
temperature ($T_{\rm eff}$), the mass of the surface He layer ($M_{\rm 
He}$), and several fixed core profiles including uniform and variable C/O 
mixtures. The best-fit model from this analysis matched the observed 
periods with a precision of $\sigma_{\rm P}=2.69$~s, but had a surface He 
layer ($M_{\rm He}\sim10^{-6}$) that was thinner than expected from 
stellar evolution theory \citep[$M_{\rm He}\sim10^{-2}$;][]{dm79}.

This pioneering work guided the development of a new model-fitting method 
based on a parallel genetic algorithm \citep{mc03} running on a 64-node 
commodity-hardware Linux cluster \citep{mn00}. Using essentially the same 
models as \citeauthor{bw94}, \citet{mnw00} had the genetic algorithm 
explore a much broader range of the three main parameters ($M_*, T_{\rm 
eff}, M_{\rm He}$) with a similar set of fixed core profiles. This 
innovative search revealed a second family of solutions with the expected 
thick surface He layers, outside of the range considered by the original 
study, which provided a considerably better match to the observations 
($\sigma_{\rm P}=1.50$~s). The results also demonstrated that the models 
were much more sensitive to the core composition than was previously 
believed. The conventional wisdom held that $g$-mode pulsations in white 
dwarf stars were primarily envelope modes. While it is true that the 
horizontal displacements in the envelope from the non-radial oscillations 
are so large that they make the motions in the core almost invisible, the 
inner 90\% of the mass actually contains the first several nodes of the 
radial eigenfunction for the pulsations observed in DBV stars.

Motivated by this newly-discovered sensitivity to the core composition, 
\citet{mwc01} extended the genetic algorithm fitting method to optimize two 
additional parameters describing a generic C/O profile. The adjustable 
central oxygen mass fraction ($X_{\rm O}$) was fixed to a constant value 
out to some fractional mass ($q$) where it then decreased linearly in mass 
to zero oxygen at 0.95~$M_r/M_*$. The optimization of these five 
parameters led to a model that matched the observations with a precision 
of $\sigma_{\rm P}=1.28$~s, representing a substantial improvement over 
the best strictly 3-parameter model ($\sigma_{\rm P}=2.30$~s). From this 
first asteroseismic measurement of $X_{\rm O}$, \citet{mwc01} attempted to 
constrain the rate of the astrophysically important $^{12}{\rm C}(\alpha, 
\gamma)^{16}{\rm O}$ nuclear reaction, extrapolating from the published 
chemical profiles of \citet{sal97}. A follow-up study by \citet{msw02} 
quantified many possible sources of systematic error in this measurement, 
but failed to identify any that were substantially larger than the 
statistical errors. However, persistent disagreement between the derived 
values of $T_{\rm eff}$ and $M_*$ and the spectroscopic estimates 
suggested room for improvement in the model.

\citet{fb02} disputed the results of \citet{mwc01}, arguing that the model 
envelopes should include a double-layered structure. If we assume that DBV 
white dwarfs are the evolutionary descendants of the hotter PG~1159 stars, 
then the surface He layer should be only $M_{\rm He}\sim10^{-6}~M_*$, 
situated above the still-uniform envelope ($M_{\rm env}\sim10^{-2}~M_*$). 
This could explain the two families of solutions for $M_{\rm He}$ found by 
\citet{mnw00} when assuming a single-layered structure. To illustrate the 
potential of double-layered envelope models, \citeauthor{fb02} attempted 
to match the pulsation periods of GD~358 with a targeted grid. Using 
several uniform core compositions, they adjusted four parameters ($M_*, 
T_{\rm eff}, M_{\rm env}, M_{\rm He}$) and found a best-fit model with 
$\sigma_{\rm P}=1.30$~s. This suggested that double-layered envelope 
models with no structure in the core could explain the observations of 
GD~358 nearly as well as the single-layered models of \citet{mwc01} which 
had an adjustable C/O profile in the core.

Around the same time, \citet{met03} identified two large sources of 
systematic error in the determination of $X_{\rm O}$. Previous fitting had 
attempted to match not only the periods, but also the forward period 
spacing ($\Delta P = P_{k+1} - P_k$) simultaneously. Ironically, this was 
intended to minimize the impact of systematic errors on the calculated 
periods, but it biased the determinations of $X_{\rm O}$ to be higher. A 
second bias in the same direction came from fixing the 
mixing-length/pressure scale height ratio too high, leading to overly 
efficient convection that modified the thermal structure of the models. By 
matching only the periods and using the ML2/$\alpha$=1.25 prescription for 
convection \citep{bc71,bea99}, the 5-parameter single-layered envelope fit 
for GD~358 improved to $\sigma_{\rm P}=1.05$~s and the derived value of 
$X_{\rm O}$ suggested a $^{12}{\rm C}(\alpha,\gamma)^{16}{\rm O}$ rate 
that was consistent with laboratory measurements \citep{ang99}.

It finally became clear that neither of these models represented an 
adequate description of the white dwarf structure when \citet{mmw03} 
identified an inherent symmetry in the way the pulsations sample the 
interior of DBV models. They found that generic perturbations caused by 
structure at specific locations in the core could not be distinguished 
from structure at corresponding locations in the envelope. This 
core-envelope symmetry could produce confusion between a C/O gradient near 
$0.5~M_r/M_*$ from prior nuclear burning, and the base of the surface He 
layer in the outer $10^{-6}~M_*$ from diffusion. However, by using 
realistic internal chemical profiles in both regions of the models, they 
determined the relative importance of the structures expected at various 
locations. This led them to conclude that the deep interior core structure 
should leave the largest imprint on the observed periods, while the base 
of the surface He layer was expected to have the smallest influence. The 
total envelope mass was less important than the core structure, but had a 
larger effect than the outermost layer.

To investigate whether the structure in the core and envelope could be 
measured simultaneously, \citet{mmk03} modified their models to include a 
parameterization of the double-layered envelopes used by \citet{fb02}. They 
performed 4-parameter fits comparable to those of \citeauthor{fb02}, as 
well as 6-parameter fits that also included an adjustable C/O profile. 
Although the addition of core structure led to a significant improvement, 
neither fit reached the level of precision attained by using 
single-layered envelope models, and the disagreement with the 
spectroscopic estimates of $T_{\rm eff}$ and $M_*$ became even worse. 
However, by fitting the periods of the model from \citet{fb02}, 
\citeauthor{mmk03} discovered a systematic offset in the temperature 
scales that could be attributed to the different radiative opacities used 
in the two models \citep{fb94}.

\begin{table*}
\centering
\begin{minipage}{100mm}
\caption{New model-fitting results for CBS~114.\label{tab1}}
\begin{tabular}{@{}lccccccc@{}}
\hline
$N_p$ & $T_{\rm eff}$~(K) & $M_*/M_\odot$ & $\log(M_{\rm env}/M_*)$ & 
$\log(M_{\rm He}/M_*)$ & $X_{\rm O}$ & $q$ & $\sigma_{\rm P}$~(s) \\ 
\hline
3 & 25,300 & 0.675 & $-$2.80 & $\cdots$ & $\cdots$ & $\cdots$ & 2.44 \\
4 & 25,800 & 0.630 & $-$2.42 & $-$5.96  & $\cdots$ & $\cdots$ & 2.33 \\
5 & 23,800 & 0.690 & $-$2.85 & $\cdots$ &   0.93   &   0.47   & 1.81 \\
6\footnote{Generic C/O profile from \citet{mwc01}} 
  & 25,200 & 0.625 & $-$2.40 & $-$5.94  &   0.91   &   0.42   & 1.51 \\
6\footnote{Physically motivated C/O profile based on \citet{sal97}} 
  & 24,900 & 0.640 & $-$2.48 & $-$5.94  &   0.71   &   0.38   & 1.27 \\
\hline
\end{tabular}
\vspace*{-18pt}
\end{minipage}
\end{table*}

Confronted with so many fitting results for one set of observations---some 
of them from models with very different interior structures---a casual 
observer might lose all confidence in our ability to do asteroseismology. 
Ultimately this is a symptom of the fact that we are confined to forward 
modeling: we adopt a structural model and see how closely it can match the 
observations. We can inform this choice with theoretical expectations, but 
we will always find an optimal model---even when the model is incomplete. 
The results from an incomplete model are not meaningless; they provide 
clues that help us identify the missing ingredients. The two families of 
solutions for the surface He layer in single-layered models told us that 
the real star showed evidence of composition transition zones at two 
locations in the envelope. The systematic differences between the derived 
masses and effective temperatures compared to spectroscopic estimates 
hinted that we needed to update the constitutive physics. How closely we 
can match the observations with each of the incomplete models gives us 
insight about which structural ingredients are the most important.


\section{NEW MODEL-FITTING RESULTS\label{SEC3}}

Recent observations of CBS~114 by \citet{met05} finally give us an 
opportunity to evaluate the models reliably on a second DBV white dwarf. 
If we fit these new data using each of the models that have previously 
been applied to GD~358---but keeping the model physics fixed---we can 
isolate the impact of various interior structures on the quality of the 
match to the observations. This exercise may be easier for CBS~114 
since the intrinsic core-envelope symmetry will affect slightly different 
regions of the model, so we might have less difficulty distinguishing the 
signatures of the surface He layer and the deep C/O profile. It should 
also give us a clearer understanding of what is actually being measured by 
each of the structurally incomplete models, and how the other parameters 
are affected by the incompleteness.

All of the modeling discussed below was performed with the WD-40 code 
originally described by \citet{mnw00}, with several important updates to 
the constitutive physics. The equation of state (EOS) for the cores and 
the envelopes come from \citet{lv75} and \citet{fgv77}, respectively. To 
describe the composition transition zones in the envelope, we use an 
adaptation of the method of \citet{af80} for single-layered models, or a 
parameterization of the time-dependent diffusion results of \citet{fb02} 
for double-layered models, as described by \citet{mmk03}. The C/O profiles 
in the core are set either by the generic parameterization developed by 
\citet{mwc01}, or by a new version that uses an adjustable profile based on 
the calculations of \citet{sal97}. The models use the updated OPAL opacity 
tables \citep{ir96}, neutrino rates from \citet{ito96}, and the 
mixing-length prescription of \citet{bc71} with ML2/$\alpha$=1.25 
convective efficiency as suggested by \citet{bea99}.

The new set of optimal models for CBS~114 are listed in Table~\ref{tab1}. 
In each case, a model-fitting method based on a parallel genetic algorithm 
was used to search a broad range for each parameter and optimize the match 
between the observed and calculated periods \citep{mc03}. The search range 
for each parameter and the details of the model-fitting procedure have 
been described by \citet{mnw00} for $N_p=3$, \citet{mwc01} for $N_p=5$, and 
\citet{mmk03} for $N_p=4$ and $N_p=6$. 

\begin{figure}
\centering
\includegraphics[width=\columnwidth]{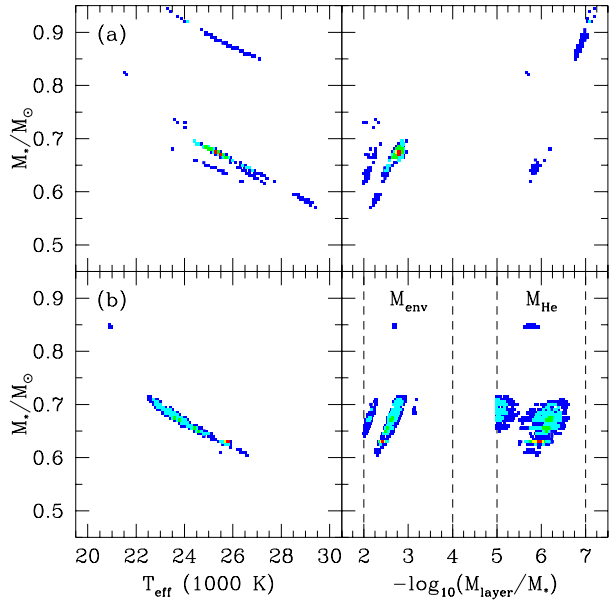}
\caption[f1.eps]{Front and side views of the search range for: {\bf a} the 
single-layered envelope ($N_p=3$) fit, and {\bf b} the double-layered 
envelope ($N_p=4$) fit to CBS~114. Models that produce residuals within
1$\sigma$ (red), 3$\sigma$ (yellow), 10$\sigma$ (green), 25$\sigma$ 
(cyan), and 40$\sigma$ (blue) of the optimum are shown.\label{fig1}}
\end{figure}

Beyond the basic stellar parameters ($T_{\rm eff}, M_*$), the simplest of 
the models ($N_p=3$) includes an adjustable mass for a single surface 
layer situated above a pure C core. The derived value of $T_{\rm eff}$ 
falls comfortably within the range determined from spectroscopy 
\citep[$T_{\rm eff}=23,300$-26,200~K, $\log g=7.98$-8.00;][]{bea99}, but 
the derived mass ($\log g \sim 8.12$) is significantly above the 
spectroscopic estimate. Despite the many simplifications inherent in this 
model, the derived envelope mass is in the range expected from stellar 
evolution theory \citep{dm79}. As with GD~358, a second family of models 
with envelope masses near $10^{-6}~M_*$ produce better than average fits 
to the observations of CBS~114, suggesting that double-layered models 
might be more appropriate (see Fig.~\ref{fig1}).

The optimal parameters for the double-layered envelope model ($N_p=4$) 
include values of $T_{\rm eff}$ and $M_*$ that are both consistent with 
the spectroscopic estimates, but the slight improvement to the fit is not 
significant. According to the Bayes Information Criterion \citep[BIC; 
see][]{mmw01}, the addition of one adjustable parameter to the fit should 
reduce the residuals from $\sigma_{\rm P}=2.44$~s for $N_p=3$ to 
$\sigma_{\rm P}=2.19$~s for $N_p=4$. The actual fit yields $\sigma_{\rm 
P}=2.33$~s for $N_p=4$, which is statistically worse than the 
single-layered fit. In the absence of core structure, single-layered 
envelope models provide a statistically better match to the observations 
of CBS~114 than double-layered envelope models. Similar results were 
obtained for GD~358, where the double-layered fit of \citet[][$\sigma_{\rm 
P}=2.17$~s]{mmk03} was also statistically worse than the single-layered 
fit of \citet[][$\sigma_{\rm P}=2.30$~s]{mnw00} for pure C cores. Even so, 
the derived values of $M_{\rm env}$ and $M_{\rm He}$ from both stars 
qualitatively agree with the expectations of diffusion theory 
\citep{met05}, suggesting that the double-layered fits might be measuring 
real structure even if the models are incomplete.

By contrast, the addition of an adjustable C/O profile to single-layered 
models ($N_p=5$) leads to a dramatic improvement in the residuals. The BIC 
leads us to expect a reduction from $\sigma_{\rm P}=2.44$~s to 
$\sigma_{\rm P}=1.96$~s, just from the addition of two adjustable 
parameters. In fact, the single-layered fit with a C/O core improves the 
fit to $\sigma_{\rm P}=1.81$~s, a statistically significant reduction. 
However, the derived values of $T_{\rm eff}$ and $M_*$ do not agree nearly 
as well with spectroscopic estimates, and the value of $X_{\rm O}$ is 
difficult to reconcile with the expectations from prior nuclear burning. 
Again, similar results were found for GD~358 by \citet{mwc01}. Despite 
these systematic errors, the derived values for $M_{\rm env}$ from the two 
single-layered fits agree with each other, suggesting that thick envelopes 
are a qualitatively robust structural feature of the results.

For models that contain an adjustable C/O profile in the core, the 
inclusion of double-layered envelopes ($N_p=6$) produces a significant 
improvement in the residuals, and restores the agreement with 
spectroscopic estimates of $T_{\rm eff}$ and $M_*$. The residuals from the 
$N_p=5$ fit should decrease from $\sigma_{\rm P}=1.81$~s to $\sigma_{\rm 
P}=1.62$~s from the addition of one adjustable parameter, but the $N_p=6$ 
fit with a generic C/O profile actually decreases the residuals to 
$\sigma_{\rm P}=1.51$~s. This statistically significant improvement is 
accompanied by derived values of $M_{\rm env}$ and $M_{\rm He}$ that once 
again qualitatively agree with diffusion theory. The only remaining 
concern with this model is the implausibly large derived value of $X_{\rm 
O}$.

\begin{figure}
\centering
\includegraphics[width=\columnwidth]{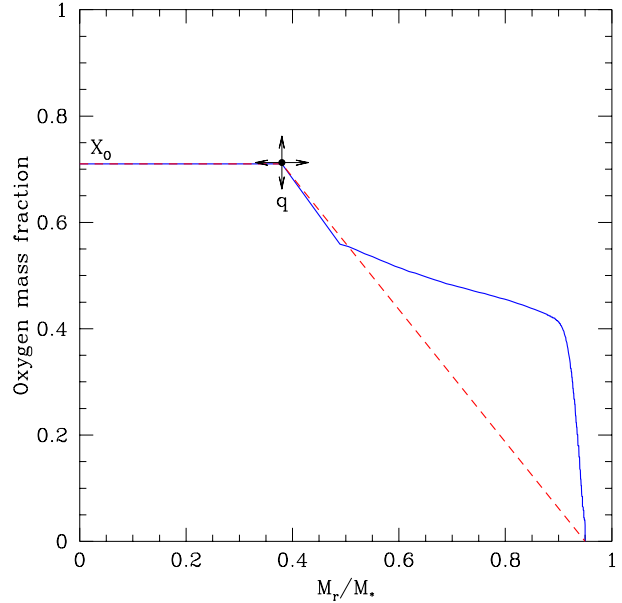}
\caption[f2.eps]{A comparison of the generic C/O profile from 
\citet[][red]{mwc01} with the physically motivated profile based on 
\citet[][blue]{sal97} used for the two $N_p=6$ fits to CBS~114. The 
plotted values of $X_{\rm O}$ and $q$ represent the final entry in 
Table~\ref{tab1}.\label{fig2}} 
\end{figure}

This may not be too surprising, considering the simplistic form of the 
generic C/O profile compared to chemical profiles generated by realistic 
stellar evolution calculations. Although a constant value of $X_{\rm O}$ 
out to some fractional mass ($q$) is a generic feature of such simulations 
\citep[see][their Fig.~2]{mmk03}, the O profile in the outer core is 
expected to be more complicated than a linear decrease. For example, the 
models of \citet{sal97} show an initially sharp drop in the O mass fraction 
near the edge of the progenitor's convective core, surrounded by a more 
gradual decline produced during He shell burning, followed by an abrupt 
drop where the temperature and pressure in the progenitor were no longer 
sufficient to sustain shell burning. A new implementation of the 
adjustable C/O profiles continues to keep $X_{\rm O}$ constant out to some 
fractional mass $q$, but then falls to zero oxygen at 0.95~$M_r/M_*$ with 
a more physically motivated shape. For $X_{\rm O}=0.76$ and $q=0.46$, the 
profile closely resembles the results of \citet[][their Fig.~3]{sal97}, 
and for any other values of $X_{\rm O}$ and $q$ the profile is scaled to 
maintain a similar shape between $q$ and 0.95~$M_r/M_*$ (see 
Fig.~\ref{fig2}).

By incorporating these physically motivated C/O profiles into 
double-layered envelope models, the optimal parameters for CBS~114 lead to 
an interior structure that agrees with both diffusion theory and the 
expected nuclear burning history. Without adding any adjustable 
parameters, this modification to the C/O profile in the outer core reduces 
the residuals from $\sigma_{\rm P}=1.51$~s to $\sigma_{\rm P}=1.27$~s, an 
improvement comparable to what would be expected from 1.5 additional free 
parameters. The derived values of $T_{\rm eff}$ and $M_*$ are in 
reasonable agreement with spectroscopic estimates, the structure of the 
envelope is approximately as expected from diffusion theory, and the core 
C/O profile implies a rate for the $^{12}{\rm C}(\alpha,\gamma)^{16}{\rm 
O}$ reaction that agrees with recent laboratory measurements. We should 
ultimately be able to refine theories of diffusion and nuclear burning to 
accommodate these new observational constraints.


\section{DISCUSSION\label{SEC4}}

The past five years have seen unprecedented progress in the development of 
white dwarf interior structure models, and in our ability to fit them to 
the available observations. This rapid improvement in our understanding 
has been most evident for the DBV white dwarfs, where the physical 
conditions are ripe for asteroseismic investigation. Our models for these 
stars have evolved from a simple parameterization of a single surface He 
layer above a pure C core to the double-layered envelope structure 
expected from diffusion calculations surrounding a physically motivated 
C/O profile in the core. We have finally been able to demonstrate that 
these structural ingredients, when taken together, lead to significant 
improvements in our ability to match the observations of CBS~114.

It is now clear that the detailed core C/O profile is the most important 
feature for quantitative asteroseismology of DBV white dwarfs. The most 
dramatic improvements in our ability to match the observations came from 
the addition of C/O cores, even when using relatively crude approximations 
for the structure in the envelope. The generic C/O profiles developed by 
\citet{mwc01} seemed to reproduce the most important asteroseismic 
features, but the physically motivated profiles introduced here appear to 
be necessary for a reliable determination of the composition deep in the 
core. Disentangling the roles of the $^{12}{\rm C}(\alpha,\gamma)^{16}{\rm 
O}$ reaction and mixing processes such as convective overshoot should be 
possible using the derived value of $q$ \citep{str03}. New simulations of 
white dwarf interior chemical profiles that attempt to match the 
asteroseismic C/O profile will help to calibrate these processes.

Without a C/O profile, the double-layered envelope structure expected from 
diffusion actually diminishes the ability of the models to match the 
observations, compared to simpler envelopes. We can now understand this to 
be a natural outcome of the relative importance of various regions of the 
interior: C/O profiles leave the largest imprint on the pulsation periods, 
followed by the base of the envelope ($M_{\rm env}$) and the surface layer 
($M_{\rm He}$). In the absence of core structure, models that maximize the 
imprint of $M_{\rm env}$ provide better fits. Since double-layered models 
effectively split the He transition zone between two separate locations in 
the envelope, the imprint of $M_{\rm env}$ is reduced relative to 
single-layered models. With the C/O core included, the fine tuning made 
possible by the double-layered models produces a significant improvement 
and leads to a credible envelope structure. Future time-dependent 
diffusion calculations should attempt to reproduce these asteroseismic 
measurements.

We can instill greater confidence is these results by applying our 
physically motivated structural models to additional white dwarf stars. 
Imposing the requirement that our model must be able to explain two 
independent sets of observations---from stars operating under slightly 
different physical conditions---is a very powerful constraint. 
Unfortunately, GD~358 may not be the best choice for such an experiment 
because the intrinsic core-envelope symmetry in this case appears to 
create an intractable ambiguity between the locations of the surface He 
layer and the core C/O profile, even when using physically motivated 
shapes for the chemical transitions. A better candidate might be found 
among DBV stars that are hotter than CBS~114 (e.g. EC~20058$-$5234) or 
cooler than GD~358 (e.g. PG~1456$+$103). Many new DBV stars have recently 
been identified in the Sloan Digital Sky Survey \citep{nit05}, but most of 
them are considerably fainter than the previously known sample. Extensive 
follow-up observations to identify the most promising multi-mode pulsators 
will require 2-m class telescopes equipped with time-series optimized 
frame-transfer CCD cameras \citep{nm04}. With the computational tools in 
place, we should quickly be able to refine our understanding of the forces 
that shape white dwarf interiors.


\section*{ACKNOWLEDGMENTS}

I would like to thank Mike Montgomery, Steve Kawaler, and Don Winget for 
helping me decode the clues that led us to improve our models over time.
This research was supported by the National Science Foundation through an 
Astronomy \& Astrophysics Postdoctoral Fellowship under award AST-0401441. 
Computational resources were provided by White Dwarf Research Corporation.




\begin{thebibliography}{}

\bibitem[Althaus \& C{\' o}rsico(2004)]{ac04} Althaus, L.~G., \& C{\' 
o}rsico, A.~H.\ 2004, \aap, 417, 1115

\bibitem[Angulo et al.(1999)]{ang99} Angulo, C.~et al.\ 1999, Nuclear
Physics A, 656, 3

\bibitem[Arcoragi \& Fontaine(1980)]{af80} Arcoragi, J. \& Fontaine, G.
1980, \apj, 242, 1208

\bibitem[Beauchamp et al.(1999)]{bea99} Beauchamp, A. et al.\ 1999, \apj, 
516, 887
 
\bibitem[B\"ohm \& Cassinelli(1971)]{bc71} B\"ohm, K. H. \& Cassinelli, J.
1971, \aap, 12, 21

\bibitem[Bradley et al.(1993)]{bww93} Bradley, P. A., Winget, D. E. \& 
Wood, M. A. 1993, \apj, 406, 661
 
\bibitem[Bradley \& Winget(1994)]{bw94} Bradley, P.~A., \& Winget, D.~E.\ 
1994, \apj, 430, 850

\bibitem[Brassard et al.(1992)]{bra92} Brassard, P., Fontaine, G., 
Wesemael, F. \& Hansen, C. J. 1992, \apjs, 80, 369

\bibitem[Dantona \& Mazzitelli(1979)]{dm79} Dantona, F., \& Mazzitelli, 
I.\ 1979, \aap, 74, 161
 
\bibitem[Dehner \& Kawaler(1995)]{dk95} Dehner, B.~T., \& Kawaler, S.~D.\ 
1995, \apjl, 445, L141
 
\bibitem[Fontaine \& Brassard(1994)]{fb94} Fontaine, G. \& Brassard, P.
1994, IAU Coll., 147, 347

\bibitem[Fontaine \& Brassard(2002)]{fb02} Fontaine, G., \& Brassard, P.\ 
2002, \apjl, 581, L33

\bibitem[Fontaine et al.(1977)]{fgv77}Fontaine, G., Graboske, H. C., Jr., 
\& Van Horn, H. M. 1977, \apjs, 35, 293
 
\bibitem[Handler et al.(2002)]{hmw02} Handler, G., Metcalfe, T.~S., \& 
Wood, M.~A.\ 2002, \mnras, 335, 698
 
\bibitem[Iben et al.(1983)]{ibe83} Iben, I. et al.\ 1983, \apj, 264, 605

\bibitem[Iglesias \& Rogers(1996)]{ir96} Iglesias, C. A., \& Rogers, F. J.
1996, \apj, 464, 943
 
\bibitem[Itoh et al.(1996)]{ito96} Itoh, N. et al.\ 1996, \apjs, 102, 411

\bibitem[Kawaler(1990)]{kaw90} Kawaler, S. D. 1990, ASP Conf. Ser. 11:
Confrontation Between Stellar Pulsation and Evolution, 494
 
\bibitem[Kawaler \& Weiss(1990)]{kw90} Kawaler, S. \& Weiss, P. 1990, in
Proc. Oji International Seminar, Progress of Seismology of the Sun and
Stars, ed. Y. Osaki \& H. Shibahashi (Berlin: Springer), 431 

\bibitem[Kepler et al.(2003)]{kep03} Kepler, S.~O., et al.\ 2003, \aap, 
401, 639
 
\bibitem[Kotak et al.(2003)]{kot03} Kotak, R., van Kerkwijk, M.~H., 
Clemens, J.~C., \& Koester, D.\ 2003, \aap, 397, 1043

\bibitem[Lamb \& Van Horn(1975)]{lv75} Lamb, D. Q. \& Van Horn, H. M.
1975, \apj, 200, 306

\bibitem[Metcalfe \& Nather(2000)]{mn00} Metcalfe, T.~S., \& Nather, 
R.~E.\ 2000, Baltic Astronomy, 9, 479

\bibitem[Metcalfe et al.(2000)]{mnw00} Metcalfe, T.~S., Nather, R.~E., \& 
Winget, D.~E.\ 2000, \apj, 545, 974

\bibitem[Metcalfe et al.(2001)]{mwc01} Metcalfe, T.~S., Winget, D.~E., \& 
Charbonneau, P.\ 2001, \apj, 557, 1021
 
\bibitem[Metcalfe et al.(2002)]{msw02} Metcalfe, T.~S., Salaris, M., \& 
Winget, D.~E.\ 2002, \apj, 573, 803

\bibitem[Metcalfe(2003)]{met03} Metcalfe, T.~S.\ 2003, \apjl, 587, L43

\bibitem[Metcalfe \& Charbonneau(2003)]{mc03} Metcalfe, T.~S., \& 
Charbonneau, P.\ 2003, Journal of Computational Physics, 185, 176

\bibitem[Metcalfe et al.(2003)]{mmk03} Metcalfe, T.~S., Montgomery, M.~H., 
\& Kawaler, S.~D.\ 2003, \mnras, 344, L88
 
\bibitem[Metcalfe et al.(2005)]{met05} Metcalfe, T.~S., Nather, R.~E., 
Watson, T.~K., Kim, S.-L., Park, B.-G., \& Handler, G.\ 2005, \aap, 435, 
649

\bibitem[Montgomery et al.(2001)]{mmw01} Montgomery, M. H., Metcalfe, T. 
S. \& Winget, D. E. 2001, \apjl, 548, L53

\bibitem[Montgomery et al.(2003)]{mmw03} Montgomery, M.~H., Metcalfe, 
T.~S., \& Winget, D.~E.\ 2003, \mnras, 344, 657
 
\bibitem[Nather et al.(1990)]{nat90} Nather, R.~E., Winget, D.~E., 
Clemens, J.~C., Hansen, C.~J., \& Hine, B.~P.\ 1990, \apj, 361, 309

\bibitem[Nather \& Mukadam(2004)]{nm04} Nather, R.~E., \& Mukadam, A.~S.\ 
2004, \apj, 605, 846

\bibitem[Nitta et al.(2005)]{nit05} Nitta, A., Kleinman, S.~J., 
Krzesinski, J., et~al. 2005, in Proc.~14th European Workshop on White 
Dwarfs, ASP Conf., 334, 585

\bibitem[Salaris et al.(1997)]{sal97} Salaris, M., Dominguez, I.,
Garcia-Berro, E., Hernanz, M., Isern, J. \& Mochkovitch, R. 1997, \apj,
486, 413

\bibitem[Straniero et al.(2003)]{str03} Straniero, O., Dom{\'{\i}}nguez, 
I., Imbriani, G., \& Piersanti, L.\ 2003, \apj, 583, 878
 
\bibitem[Vuille et al.(2000)]{vui00} Vuille, F., et al.\ 2000, \mnras, 
314, 689
 
\bibitem[Winget et al.(1994)]{win94} Winget, D.~E., et al.\ 1994, \apj, 
430, 839

\bibitem[Winget et al.(2004)]{win04} Winget, D.~E., Sullivan, D.~J., 
Metcalfe, T.~S., Kawaler, S.~D., \& Montgomery, M.~H.\ 2004, \apjl, 602, 
L109

\end{thebibliography}
\end{document}